\documentstyle[aas2pp4]{article}

\begin{document}
\title{White Dwarfs Near Black Holes: A New Paradigm for Type I Supernovae}

\author{J. R. Wilson}
\affil{Lawrence Livermore National Laboratory, Livermore, CA 94550}
\centerline{\it  and} 
\affil{ 
Center for Astrophysics,
Department of Physics,
University of Notre Dame,
Notre Dame, IN 46556}

\vskip .2 in
\centerline{\it  and} 

\author{G. J. Mathews}

\affil{
Center for Astrophysics, 
Department of Physics,
University of Notre Dame,
Notre Dame, IN 46556}

\date{\today}
\begin{abstract}
We present calculations indicating the possibility of  a new class of Type I supernovae.
In this new paradigm relativistic  terms enhance the self gravity of a carbon-oxygen 
white dwarf as it passes or
orbits near a black hole.  This relativistic compression can
cause the central density to exceed the threshold for 
pycnonuclear reactions so that a thermonuclear runaway ensues.  
We consider three possible environments:
1)  white dwarfs orbiting a low-mass  ($\sim 10-20 $ M$_\odot$) black hole; 2)
white dwarfs encountering a massive ($\sim 1-3\times 10^3 $ M$_\odot$)
black hole in a dense globular cluster; and 
3) white dwarfs passing a supermassive ($\sim 10^6 - 10^9 $ M$_\odot$) black hole in 
a dense galactic core.    
We estimate the rate at which
such events could occur out to a redshift  of $z \approx  1$. 
Event rates are estimated to be significantly less than the rate
of normal Type Ia supernovae for all three classes.  Nevertheless,
such events may be frequent enough
to warrant a search for this new class of supernova.  
We propose several observable signatures which might be used
to identify this type of event and speculate that such an  event
might have produced the observed  "mixed-morphology" 
Sgr A East supernova remnant in the Galactic core.     
\end{abstract}

\section{Introduction}

Type I supernovae are currently objects of intense interest.   The observed luminosity
of Type Ia supernovae (SNIa)  out to high redshift offers some of the most convincing evidence
for the acceleration of the universe due to an unknown dark energy (Garnavich et al
1998; Riess et al. 1998; Perlmutter et al. 1998).
The distance redshift relation for the SNIa  standard candle is
also an important  component in the determination of the present cosmic expansion rate
(Freedman et al. 2001).
As such, it is important to carefully scrutinize any possible
variations in the paradigm for SNIa explosions which might occur at large redshift.
In this paper we explore one such possible new mechanism for SNIa's.
 
Type I supernovae are characterized by the absence of hydrogen lines
in their spectrum.  Type Ia supernovae, in particular, are believed to result
from the accretion of material onto the surface of a carbon-oxygen white dwarf. 
If the accretion rate is sufficiently rapid,
mass accumulates until the star approaches the Chandrasekhar limit.  As the
interior density exceeds the threshold for pycnonuclear
reactions, an off-center  runaway  deflagration/detonation wave of
thermal carbon burning ignites under strongly degenerate conditions.
Since the degeneracy pressure support of the core is almost
independent of temperature, the heating of the core does not cause
the star to expand and cool.   Therefore, carbon burning rates increase until
a thermonuclear explosion eventually engulfs a significant fraction of the star.
The key to this paradigm is that the density at some time exceeds the 
threshold for thermonuclear ignition. For typical Type Ia models
the ignition density ranges from 
$\rho_{ign} \approx 2-4 \times 10^9$ g cm$^{-3}$ (at a temperature
$T \sim 5 \times 10^8$ K) with the lower range of ignition
density favored by nucleosynthesis arguments (Iwamoto et al. 1999).

 In this paper we propose a new  mechanism by which the ignition density
of a white dwarf can be achieved.
We show that relativistic enhancements of the self gravity
of a white dwarf occur whenever the star is accelerating in a background gravitational field.
As the velocity with respect to the background field approaches a significant fraction
of the speed of light (as will occur near a black hole)
these relativistic effects can cause the central density to exceed the threshold for nuclear
ignition  even for white dwarfs which are well below the Chandrasekhar mass.
Alternatively, the ignition densities might be achieved by the excitation of
tidally-induced resonance/disruption
 (Khokhlov \& Melia 1996; Rathore \& Blanford 2000) within the white dwarf,
or tidal compression by the Carter-Luminet effect (Carter \& Luminet 1982;
Laguna, Miller \& Aurek 1994).
In any of these scenarios, the result is the same, i.e. the ignition
of a thermonuclear explosion in a deep gravitational well.
The focus of the present work, however, is on the relativistic compression effect
(Wilson \& Mathews 2000; Wilson 2002).

We note that other authors have modeled hydrodynamic effects 
which occur as stars encounter a black hole.  
For example, Fryer et al. (1999) have considered the tidal disruption
of a white dwarf near a low mass black hole as a source of gamma-ray bursts.  
Their simulations were based upon Newtonian hydrodynamics which would not incorporate the
general relativistic effect described here.  Similarly, Ayal, Livio, and Piran (2000)
have simulated the tidal disruption of a normal solar type main-sequence
star near a massive ($m \sim 1000$ M$_\odot$)
black hole.  Extended main sequence stars indeed should tidally disrupt before
the relativistic effects described here can induce compression.  
The effect described here is unique
to compact objects such as white dwarfs or  neutron stars and is only apparent in a 
fully relativistic simulation.

We consider three possible environments in which such events might occur.
These are: 1)  white dwarfs orbiting a low-mass black hole (hereafter LMBHSN); 2)
white dwarfs encountering a massive  ($M > 1000$ M$_\odot$ black hole in a globular cluster 
 (hereafter MBHSN); and 3) white dwarfs passing a supermassive ($M > 10^6$ M$_\odot$ 
black hole in a dense galactic core   (hereafter SMBHSN).  
In what follows we  utilize  general relativistic hydrodynamic
numerical simulations described in Sections 2 and 3 to determine
the conditions under which the central density can be made to exceed 
the ignition threshold for these three paradigms.  In Section 4 we then
make a rough analysis  of the event rates at  which these various scenarios might
occur.  We conclude with some discussion as to how such events might
be identified observationally.

\section{White-Dwarf Compression}

It has been noted  in studies of colliding (Wilson 2002) and orbiting
(Mathews \& Wilson 2000) neutron stars that general relativistic
effects can overwhelm the stabilizing tidal forces and
cause the central density of compact objects  to increase.  
 This phenomenon was shown (Mathews, Marronetti \& Wilson  1998; 
Mathews \& Wilson 2000)
 to scale with the magnitude of the spatial component of the four velocity of the star relative
to the background field for
conditions  in  which there exists no Killing vector by  which the three-velocity
can be made to vanish.
 
Here we point out that these conditions are also met for
white dwarfs orbiting or  passing near  a black hole.  Thus,     
a similar scaling of interior density with
four velocity should be expected for the case of white dwarfs accelerating under the
influence of the gravitation potential of a massive black hole.  
We show by numerical calculations that
the increase of central density is sufficient to induce a thermonuclear runaway
and lead to a new class of Type I supernovae.  
These supernovae
are unique in that they would most likely be found in association with  
a black hole. The case of MBHSNe or SMBHSNe would also be associated with
dense globular clusters or galactic cores, respectively.  
They would also  be distinguished from the
usual Type Ia event in that they may involve a less massive white dwarf
and a rapid ignition/detonation  in their centers.

\subsection{The Model}
We analyze the induced compression in  the context
of a general relativistic treatment of the 
hydrodynamic and field evolution of a white dwarf in the vicinity of a black hole.
Although an orbiting system is inherently nonaxisymmetric, 
the essential features of this system can be modeled (with slight modifications)
while imposing axisymmetry. The implementation of axisymmetry
does not lead to significant inaccuracies, but it 
allows us to rapidly explore a broad range of
 parameter space for white-dwarf plus
black-hole systems as follows.

An  exact axisymmetric general relativistic hydrodynamic
model has been previously developed by Wilson (1979; 2002)
(see also  Hawley, Smarr \& Wilson 1984)
to describe the head-on collision of two neutron stars.
Here we have modified this model to treat the essential
features of a white-dwarf plus black-hole system.

We start with  the ADM formalism (Arnowitt, Deser \& Misner 1962)
whereby, the metric is split into space-like and time-like components
\begin{equation}
ds^2 = -(\alpha^2 - \beta_i\beta^i) dt^2 + 2 \beta_i dx^i dt + \gamma_{ij}dx^i dx^j~~,
\label{admeq}
\end{equation}
where $\alpha$ is the lapse function, $\beta^i$ is the shift vector, and
$\gamma_{i j}$ is the metric for the three-space (we use geometrized units.
$G = c = 1$).  For the present purposes,
we are concerned with either quasi-stationary orbits or passing orbits near
the point of closest approach.  In either case, the $\beta^r$ and $\beta^z$ metric
components are very small  for the systems of interest here
and can be neglected in what follows.
In a full nonaxisymmetric treatment, the
dominant contribution to the shift vector will be  from 
nonaxisymmetric motion about the rotation axis of the orbit plane.
Nevertheless, an axisymmetric  model can still be maintained
by modified hydrodynamic equations as described below.

In essence there are two effects of orbital motion on the shift vector:
In a comoving gauge  (Wilson \& Mathews 2003) the dominant contribution
is from an overall rotation of the coordinates. The explicit
inclusion of this term naturally generates the appropriate
centrifugal and Corriolis forces associated with a rotating coordinate system.   
There is also a small Lense-Thirring frame-drag
correction (Wilson, Mathews \& Marronetti 1996) for the relativistic orbit precession.
The frame-drag component can be deduced from an equation of the form,
$\nabla^2 \beta_{drag} \approx 16 \pi \alpha \phi^4 W  \rho_{WD} U$   (Wilson \& Mathews 2003).
For the systems of interest here, the  source term is small
 so that  frame-drag is
only a small correction to the orbital motion.
This is consistent with numerical results for neutron-star binaries
near their last stable orbit (Mathews \& Wilson 2000).
Hence, for our purposes the frame-drag can be neglected.  

The coordinate rotation term can be eliminated by choosing an 
instantaneously nonrotating coordinate system at the orbit periapse
and adding effective centripetal force terms to the 
equations of motion as described below.

Thus, for this application we adopt a quasi-Eulerian ($\beta_i = 0$) 
gauge (Wilson \& Mathews 2003).
We further adopt a conformally flat metric whereby
$\gamma_{ij} = \phi^4\delta_{ij}$.  This has been shown
(Cook et al. 1993; Mathews \& Wilson 2000; Wilson 2002) 
to be  sufficiently accurate for our purpose.

 Thus, in cylindrical coordinates we write an effective metric ,
\begin{equation}
 ds^2 = -\alpha dt^2 + \phi^4 [ dr^2 + dz^2 + r^2 d\theta^2]~~,
\end{equation}
where in what follows we take the $z$ coordinate to be
the  distance along the
line of centers between the black hole and the white dwarf.  The $r$ coordinate
is the radius perpendicular to the $z$  axis.
This metric leads  (Wilson, Mathews \& Marronetti 1996;  Wilson 2002;
Wilson \& Mathews 2003)
to simple Poisson-like  elliptic constraint equations which can be  
solved to find the the conformal factor and lapse function.
\begin{eqnarray}
\nabla^2{\phi} &= &-2\pi\phi^5 \biggl[DW +
E \biggl( \Gamma W - \frac{(\Gamma-1)}{ W}\biggr)
\nonumber\\
&& + \frac{1}{16\pi} K_{ij}K^{ij}\biggr]~~.
\label{rho1}
\end{eqnarray}
\begin{eqnarray}
\nabla^2(\alpha\phi)& =& 2\pi\alpha \phi^5 
\nonumber\\
& \times& \biggl[\frac{D (3W^2-2)+
E[ 3\Gamma (W^2+1)-5]}{W} 
\nonumber\\
&& + \frac{7}{16\pi} K_{ij}K^{ij}\biggr]~~.
\label{rho2}
\end{eqnarray}
 where the source term includes Lorentz contracted state variables,
$D = \rho W$, $E = \epsilon \rho W$,  with $\rho$ the local
proper mass density,  $\epsilon$ the internal energy
per unit mass, and $W$ a Lorentz-like factor,
$W = \sqrt{1 + U^iU_i}$, where $U^i$ denotes spatial components of
the four velocity.  

The quantity $\Gamma \equiv  1 + P/ \epsilon \rho$ is an equation
of state index.  
We use a standard white dwarf equation of state
(e.g. Salpeter 1961; Hamada \& Johnson 1961) to specify $\epsilon$ and/or $\Gamma$
as a function of density.  
For simplicity we only retain the dominant degenerate electron part.

In equations  (\ref{rho1})
 and (\ref{rho2}) the extrinsic curvature $K_{ij}K^{ij}$  terms result 
(Wilson \& Mathews 2003) from
derivatives of the relativistic frame drag contained in the $\beta^i$.
In close neutron-star binary systems  
it has been shown (Mathews \& Wilson 2000) 
that this term contributes insignificantly
to the metric source terms, 
($\int dV K_{ij}K^{ij}/8 \pi \sim 10^{-5} M_{Grav}$).
In what follows
these extrinsic curvature  terms can therefore be neglected.  
Including these terms would only slightly enhance
the compression effect described below.

\subsection{ Axisymmetric Hydrodynamics}

To model a white dwarf near a black hole, a centrifugal term
is added to the axisymmetric momentum equations to keep the stars apart.
Thus, the hydrodynamic equations are slightly modified from
those presented in (Wilson 2002; Wilson \& Mathews 2003)
The momentum equation becomes,
\begin{eqnarray}
{\partial S_i\over\partial t} & +& 6 S_i{\partial \ln\phi\over\partial t}
+{1\over\phi^6}{\partial\over\partial x^j}\biggl(\phi^6S_iV^j\biggr)
+\alpha{\partial P\over \partial x^i}
\nonumber \\
& +& \sigma \alpha \biggl[(1 + U^2 )  {\partial \ln{\alpha} \over \partial x^i}
 - 2 U^2 {\partial \ln\phi\over\partial x^i} \biggr]
\nonumber \\
&-& \sigma {U_\bot^2 \over z} \delta_{i z}  = 0~~,
\label{hydromom}
\end{eqnarray}
where $\sigma \equiv D + \Gamma E$, and the covariant coordinate momentum
density is $S_i = \sigma U_i$, and $U^2 \equiv U_i U^i$.

  The last term in Eq. (\ref{hydromom})
is a centrifugal acceleration.  It  accounts for 
the conserved motion along
an unspecified third coordinate perpendicular to 
$z$.  Obviously,
setting this term to zero recovers the simple 
 head on collision.   If  this term is  sufficiently large, 
it can stabilize the stars in orbit, or it can limit motion
along $z$ to a distance of closest approach.  

It is to be
noted that this component must be included in the $U_i U^i$ summations
above and in the metric fields.  For example, at closest approach,
$U^2 = U_\bot U^\bot$.    Thus, the inclusion of this term
increases the source terms [cf. Eqs.~(\ref{rho1}) and (\ref{rho2})] for
the solution of the metric fields.

The remaining hydrodynamic equations assume their usual form.
The continuity equation is,
\begin{equation}
{\partial D\over\partial t}  +   6D{\partial \ln\phi\over\partial t}
 + {1\over\phi^6}{\partial\over\partial x^j}(\phi^6DV^j) = 0~~.
\end{equation}
The equation  of internal energy evolution is,
\begin{eqnarray}
{\partial E\over\partial t}  &+&  6\Gamma E{\partial \ln\phi\over\partial t}
 + {1\over\phi^6}{\partial\over\partial x^j}(\phi^6EV^j)
\nonumber \\
&+&
  P\biggl[{\partial W\over \partial t} +
{1 \over\phi^6}{\partial\over\partial x^j}(\phi^6 W V^j)\biggr] = 0 ~~,
\end{eqnarray}

\subsection{Enhanced Self Gravity}

The term with $\sigma \alpha {\partial  \ln{\alpha}/\partial x^i}$ 
in Eq.~(\ref{hydromom}) is the
the relativistic analog of the  Newtonian gravitational force.
The general relativistic
condition of hydrostatic equilibrium for
each star can be inferred by taking the time-stationary limit 
of the momentum equation (\ref{hydromom}), and transforming to
instantaneously comoving  
three space coordinates with a grid velocity  $x^i_g = x^i - V_g^j dt$
such that the three velocity $V^j - V_g^j = 0$.  This gives 
\begin{equation}
{\partial P \over  \partial x^i} = -\sigma 
\biggl( {\partial \ln{\alpha} \over \partial x^i} 
 + U^2 \biggr[ {\partial \ln{\alpha} \over \partial x^i} -
  2 {\partial \ln{\phi} \over \partial x^i}\biggr]  \biggr)~~.
\label{selfg}
\end{equation}
Note, that in the weak-field limit, $\phi^2 \approx \alpha^{-1}$
which reduces Eq. (\ref{selfg}) to
\begin{equation}
{\partial P \over  \partial x^i} \approx -\sigma
{\partial \ln{\alpha} \over \partial x^i}
 \biggl(1  +2  U^2 \biggr)~~,
\label{selfg2}
\end{equation}
from which it is manifestly apparent that the terms in the square brackets of
Eq. (\ref{selfg}) do not cancel.
The enhanced self gravity derives from the additional
$U^2$ dependent terms of Eqs. (\ref{selfg}) and (\ref{selfg2}).

\section{Model Calculations}

As noted above, we consider several possibilities in this study.
One is a white dwarf in a bound orbit around a low-mass
black hole.  This situation could result, for example,
 from the evolution of a binary
in which one member was a massive progenitor which formed a black hole.
Alternatively, a white dwarf might be captured by a black hole through
three-body interactions in a dense 
cluster.   In either case,
the white dwarf orbit would decay by gravitational
radiation until it approaches high orbital velocity.
Another possibility is that of a white dwarf 
randomly passing a massive $_\sim^>  1000$ M$_\odot$
black hole in a dense globular cluster or galactic core.  

To start each calculation an initial distribution of density and
internal energy  is taken
(e.g.~a white dwarf in hydrostatic equilibrium) and with initial velocities
corresponding to either nonradial infall from infinity, or of an amount required to
stabilize the orbit, i.e.~an angular momentum is chosen to
present a centrifugal barrier. In the case of a simple bound circular orbit,
this term was simply adjusted to keep the white dwarf at a fixed distance
from the black hole.  For the case of passing orbits, this term was chosen to
produce a particular distance of closest approach.
The stars were then evolved and the central density monitored
until the white dwarf was either tidally disrupted, attained the critical
density for thermonuclear burn, or passed the distance
of closest approach.
 
For our purposes, a thermonuclear runaway is conservatively expected
to initiate once the density exceeds $3 \times 10^9$ g cm$^{-3}$.
Table 1 summarizes key properties of white dwarfs  of various mass  in bound circular orbits
 around low-mass black holes. Tables 2 and  3 summarizes some calculations
of passing orbits of massive and supermassive black holes, respectively,
 at the points of closest approach. For illustration, column 3 in these tables
gives $\rho_c^\infty$ the central density for isolated white dwarfs of the indicated mass.

\begin{deluxetable}{lccccc}
\tablenum{1}
\tablewidth{0pt}
\tablecaption{ White Dwarfs Orbiting a Black Hole }
\tablehead{
\colhead{M$_{WD}$  (M$_\odot$)}  & \colhead{M$_{BH}$ (M$_\odot$)}  &
\colhead{ $\rho_c^\infty$ ($10^9$ g cm$^{-3}$)} &
\colhead{ $\rho_c$ ($10^9$ g cm$^{-3}$)} &\colhead{ $z_{min}$ (km)} &\colhead{ $U^2$}
}
\startdata
\tableline
1.20   & 10   & $0.12  $ & $1.2 $ & $4.82 \times 10^3$ & 0.0031  \\
1.20   & 20   & $0.12  $ & $1.6 $ & $6.52 \times 10^3$ & 0.0045 \\ 
1.30   & 10   & $0.37  $ & $2.9 $ & $4.21 \times 10^3$ & 0.0035  \\ 
1.30   & 20   & $0.37  $ & $3.0 $ & $5.77 \times 10^3$ & 0.0051 \\ 
\\ 
\tableline
\enddata
\end{deluxetable}
\begin{deluxetable}{lccccc}
\tablenum{2}
\tablewidth{0pt}
\tablecaption{ White Dwarfs  Passing a Massive Black Hole}
\tablehead{
\colhead{M$_{WD}$  (M$_\odot$)}  & \colhead{M$_{BH}$ (M$_\odot$)}  &
\colhead{ $\rho_c^\infty$ ($10^9$ g cm$^{-3}$)} &
\colhead{ $\rho_c$ ($10^9$ g cm$^{-3}$)} &\colhead{ $z_{min}$ (km)} &\colhead{ $U^2$}
}
\startdata
\tableline
0.60   & 1000 & $0.0032 $ & $3.4 $ & $0.120 \times 10^5$ & 0.248   \\ 
0.60   & 1500 & $0.0032 $ & $3.2 $ & $0.193 \times 10^5$ & 0.231   \\ 
0.60   & 3000 & $0.0032 $ & $3.9 $ & $0.393 \times 10^5$ & 0.226   \\ 
0.80   & 3000 & $0.0091 $ & $3.6 $ & $0.706 \times 10^5$ & 0.126  \\ 
1.00   & 3000 & $0.029  $ & $3.1 $ & $1.24  \times 10^5$ & 0.0715 \\ 
1.20   & 3000 & $0.12   $ & $3.1 $ & $2.81  \times 10^5$ & 0.0316 \\ 
1.30   & 3000 & $0.37   $ & $3.2 $ & $5.50  \times 10^5$ & 0.0160 \\ 
1.40   & 3000 & $1.9    $ & $3.2 $ & $33.0  \times 10^5$ & 0.0027  \\ 
\\ 
\tableline
\enddata
\end{deluxetable}
\begin{deluxetable}{lccccc}
\tablenum{3}
\tablewidth{0pt}
\tablecaption{White Dwarfs Passing a Supermassive Black Hole}
\tablehead{
\colhead{M$_{WD}$  (M$_\odot$)}  & \colhead{M$_{BH}$ (M$_\odot$)}  &
\colhead{ $\rho_c^\infty$ ($10^9$ g cm$^{-3}$)} &
\colhead{ $\rho_c$ ($10^9$ g cm$^{-3}$)} &\colhead{ $z_{min}$ (km)} &\colhead{ $U^2$}
}
\startdata
\tableline
0.60 & 10$^9$ &  $0.0032 $  & $3.06 $ & $1.5 \times 10^{10}$ & 0.191   \\ 
0.70 & 10$^9$ &  $0.0056 $  & $3.29 $ & $2.0 \times 10^{10}$ & 0.151   \\ 
0.80 & 10$^9$ &  $0.0091 $  & $3.05 $ & $2.5 \times 10^{10}$ & 0.118   \\ 
0.90 & 10$^9$ &  $0.010  $  & $2.95 $ & $3.3 \times 10^{10}$ & 0.0905  \\ 
1.00 & 10$^9$ &  $0.029  $  & $3.13 $ & $4.3 \times 10^{10}$ & 0.0686  \\ 
\\ 
\tableline
\enddata
\end{deluxetable}

Most of these table entries have been selected because they represent systems
which have just achieved the ignition density in their centers
at the distance of closest approach.  The two entries
of a 1.2 M$_\odot$ white dwarf orbiting a low mass black hole, however, 
are just below our adopted ignition density.  All of the stars in Table 1
have large tidal distortion
even though their central density is quite high.

\begin{figure}
\plotone{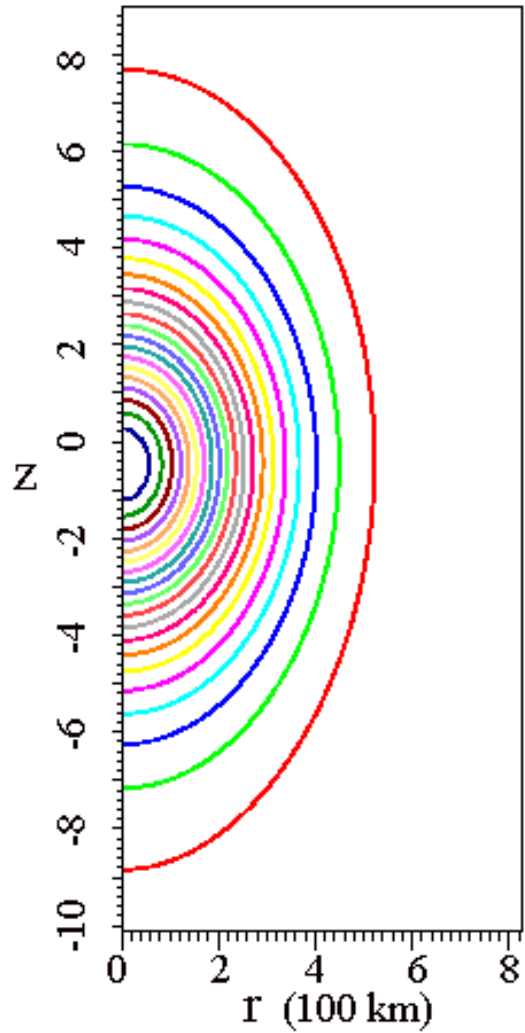}
\caption[]{
Contours of constant density  in the $r$ vs. $z$ plane
for a 0.6 M$_\odot$ white dwarf
approaching a 1000 M$_\odot$ black hole.
The star is near closest approach,
centered at $ z \approx 1.4 \times 10^4$ km from the black hole
which is located below the figure.
This star is tidally distorted to an aspect ratio of 
$(z,r) \approx  (900,500)$ km.  Nevertheless, the radius has contracted
by a factor of 10 from that of an isolated star.
The central density has correspondingly increased 
 by more than a factor of 1000 from 
($3.2 \times 10^6$) to  ($3.3 \times 10^9$ g cm$^{-3}$)
which is above the pycnonuclear ignition threshold.
}
\label{wdfig1}
\end{figure}

\begin{figure}
\plotone{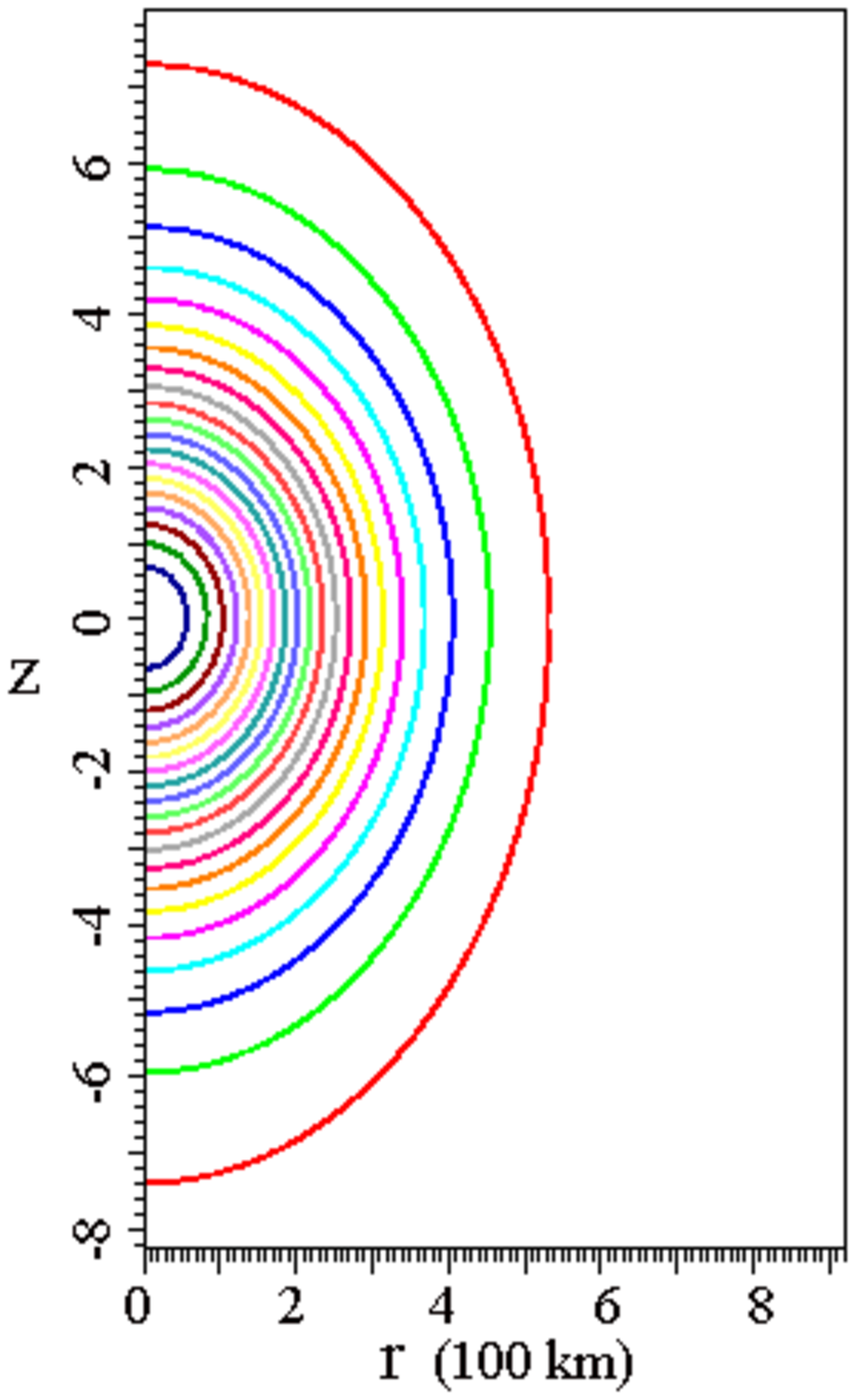}
\caption[]{
Contours of constant density in the $r$ vs. $z$ plane
 for a 0.6 M$_\odot$ white dwarf
near closest approach ($ z \approx 2 \times 10^4$ km) to a 1500 M$_\odot$ black hole.  
}
\label{wdfig2}
\end{figure}

\begin{figure}
\plotone{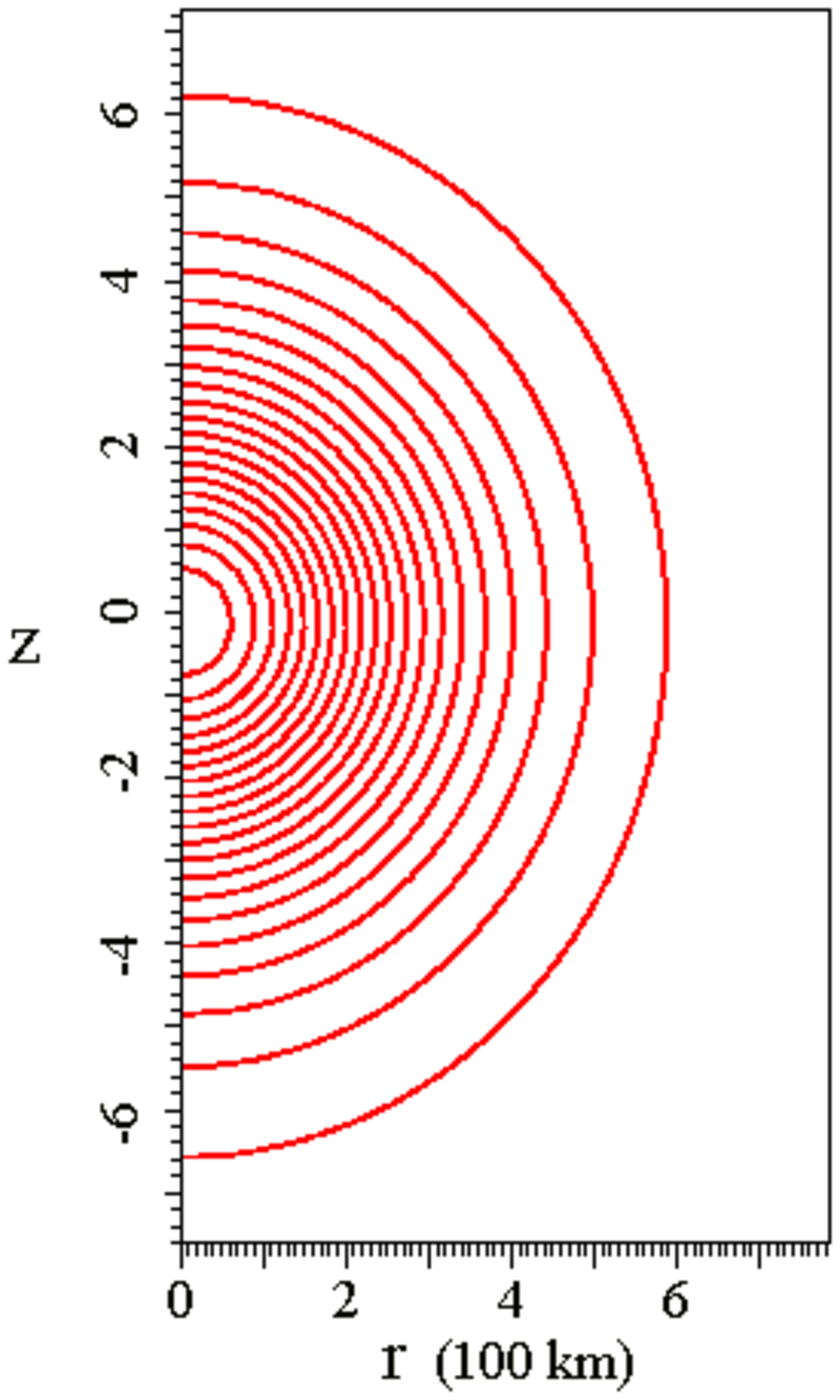}
\caption[]{
Contours of constant density in the $r$ vs. $z$ plane
for a 0.6 M$_\odot$ white dwarf
near closest approach ($ z \approx 4 \times 10^4$ km) to a 3000 M$_\odot$ black hole.  
}
\label{wdfig3}
\end{figure}

Figures \ref{wdfig1}  to  \ref{wdfig3}  illustrate the
competition between tidal distortion and relativistic compression.
These figures are for a typical 0.6 M$_\odot$ white dwarf
passing  near various massive black holes.
Note, that the radius of an isolated 0.6 M$_\odot$ white dwarf is
$\sim 9000$ km.  Hence, these stars are significantly compressed.
The black hole resides at $\sim 10^4$ km below these figures
as noted in the captions.
All three cases have been chosen because 
the white dwarf just achieves the ignition threshold
in its center near the point of closest approach to the
black hole. These therefore represent the maximum impact parameter for which
explosion can occur.  

Figure \ref{wdfig1} is for a white dwarf
near a 1000 M$_\odot$ black hole.  
In this case, the white dwarf is substantially tidally distorted.
Nevertheless it has compressed (by more than a factor of 1000 in density
and a factor of 10 in radius) so that the
center reaches the ignition threshold.
Figure \ref{wdfig2} is for a 1500 M$_\odot$ black hole.  
It  shows less tidal distortion
since it reaches the central ignition density farther from the star. 
Similarly,  Figure \ref{wdfig3} for a white dwarf
even farther from a 3000 M$_\odot$ black hole exhibits almost no tidal 
distortion at all.
These figures clearly illustrate that for such systems, the tidal forces
(which tend to decrease the central density)
are not strong enough to prevent an increase in central density as the white dwarf
approaches the black hole.

From these simulations the following conclusions can be made regarding this paradigm:
\begin{itemize}
\item Substantial increases in central density can occur as a star passes close to
a black hole.  For example, the central density of  a  $0.6$ M$_\odot$ white dwarf
in isolation is $3.2 \times 10^6$ g cm$^{-3}$.  This density is increased by nearly
a factor of $10^3$ when passing a massive black hole.

\item The ignition density can be achieved for smaller velocities for white dwarfs
that are closer to the Chandrasekhar mass.

\item For low-mass black holes, pycnonuclear ignition requires the orbit to be relatively
close  ($\sim {\rm few} \times  10^3 $ km), i.e comparable to
the radius of the white dwarf.  Also, the white dwarf must be in the high
end of the observed mass distribution, $M_{WD} > 1.2$ M$_\odot$.  However,
For a black hole of $m_{BH} \ge 1000$ M$_\odot$ 
sufficient compression can occur  even for a 0.6 M$_\odot$ white dwarf, and
ignition can occur much further from the black hole where tidal
effects are minimal (cf. Fig. \ref{wdfig3}).
\end{itemize}

Hence, we conclude that the most likely environment for this effect
is in a dense galactic core or globular cluster in which there is a
good chance that a white dwarf can pass relatively close to a massive
or supermassive black hole.

\section{ Event  Rates}

\subsection{Normal Type Ia Supernovae}
It is important to have an estimation of the rates of such new 
events relative to those of normal SNIa's.  The observable event rate for normal
SNIa explosions out to a redshift of $z \approx 1$ is rather significant
\begin{equation}
\dot N_{SNI} = N_{gal} \dot R_{Ia}  
\approx 10^9 \times 10^{-2} \approx  10^7~~{\rm yr}^{-1}~~,
\end{equation}
where $N_{gal}$ is the number of galaxies out to $ z \approx 1$, and
$R_{Ia}$ is the average rate per galaxy
at which carbon-oxygen white dwarfs accrete enough material to exceed the
critical interior density for the ignition of an explosion.
This rate is observed to be about 1 per 
century per galaxy (Kobayashi et al. 2000).
It is instructive, however, to review how one estimates this rate
theoretically:

    If we adopt the current view that
 Type Ia  supernovae arise from the binary accretion of
material onto a white dwarf.  There are two scenarios which
have been debated in the literature (Nomoto et al. 1994;
Kobayashi et al 2000).  
The most widely adopted from a theoretical modeling point of view
(and the one that will be adopted here) is that of   
 a single degenerate white dwarf which accretes hydrogen
via mass transfer from a normal red giant or main sequence companion. If the mass
transfer is sufficiently rapid, the interior density rises to the ignition
density before material can be ejected via normal nova eruptions.
A second possibility (Iben \& Tutukov 1984; 1985)
is that of a doubly degenerate carbon-oxygen white-dwarf binary.
The eventual  merger
of the two white dwarfs leads to a single system of order the
Chandrasekhar mass which can ignite and explode. 

 The relative rates of either of these paradigms 
can be deduced from the relative probability of having
a close interacting binary.  For this
we utilize a slightly
modified version (Mathews, Bazan \& Cowan 1992)
 of an expression  first proposed by Greggio \&  Renzini (1983).
\begin{eqnarray}
R_{Ia}(t) &=& \int_{M_{l}}^{M_{h}} dM_{B} \phi (M_{B})
\int_{q_{l}}^{q_h}
dq f(q) 
\nonumber 
\\
&\times&
\int_{a_{l}(t,M_{B},q)}^{a_{h}(t-\tau,M_{B},q)} da
P(a) \psi (t-\tau)~,
\label{rsn1a}
\end{eqnarray} 
where $\phi(M_{B})$ is the initial mass function (IMF) 
for a binary of total mass, $M_{B}$,
$f(q)$ is the distribution function for the ratio of the secondary to
primary mass, $q = m_{2}/m_{1}$, $P(a)$ is the distribution of
initial separations, $a$, and $\psi(t-\tau)$ is the star formation
rate corrected for the evolutionary time  $\tau$ for the secondary star
to evolve to a white dwarf and for the binary to evolve to a supernova.
As in (Mathews, Bazan \& Cowan 1992) we take 
$f(q) \propto q^{x}$ with $x < 2$, since distributions with exponents $\ge
2$ give incorrect ratios of single lined to double lined spectroscopic binary
systems (Kraicheva {\it et al.} 1979).  This favors mass-symmetric
systems. The initial separation distribution is
taken by similar arguments (Tutukov \&  Yungelson 1979) to be $P(a) \propto
a^{-1}$, which favors close systems.

    The upper and lower limits to the mass integral are  simply the largest
 and smallest total binary mass, respectively,  allowed by the
scenario and  evolutionary
time scale.   For a single degenerate white dwarf SNIa, one typically has,
$1 \le m_2 \le 3$ M$_\odot$ and $3 \le m_1 \le 8$ M$_\odot$ as the carbon-oxygen white dwarf
progenitor.  This implies, $M_l = 4$, $M_h  = 11$ and $q_l = 1/8$, $q_h = 1$.
 The limits on the separation
distance are determined by the presumed scenario for Type Ia supernovae as
discussed above. For our purposes, the lower limit $a_l \sim 10$ AU is determined 
(Mathews et al. 1992) by the requirement
that the stars not be too close during the common-envelope 
asymptotic-giant-branch phase (Iben \& Tutukov 1984; 1985) 
of the white dwarf progenitor
to prevent the formation of a CO white dwarf.

\subsection{White Dwarfs Orbiting a Black Hole}

One might think that a white dwarf orbiting a common low-mass
black hole could be the most likely candidate for this event.
After all, a  number of candidate binary systems have been identified
(Cowley 1992) which probably contain  a black holes as one member.
Indeed, mass transfer from a close companion star is the primary
means to identify a black hole candidate.  Hence, it seems quite
likely that a number of close binary systems exist in which the higher-mass
progenitor star has evolved to a black hole while the lower-mass member
is now a white dwarf.  In this case, the inspiraling of the
white-dwarf companion is an inevitable outcome and the possibility of
a compression-induced explosion needs to be considered.
Here, we  estimate the rate of such events relative to
normal SNIa events.  We find that this is a possible, though rare,
phenomenon.

 The relative rate of exploding
white-dwarf, black-hole binaries
compared to that of normal Type I supernovae can be deduced from Eq.~(\ref{rsn1a}).
Ignoring the logarithmic dependence on relative separation,
 and assuming instantaneous recycling
along with a Salpeter IMF, the ratio of
rates becomes:
\begin{eqnarray}
{R_{LMBHSN} \over R_{SN1a}} &\approx&
{[M_l^{-1.3} - M_h^{-1.3}]_{LMBHSN} \over [M_l^{-1.3} - M_h^{-1.3}]_{SN1a}}
\nonumber \\
&\times &{[q_{max}^{x+1} - q_{min}^{x+1}]_{LMBHSN} \over [q_{max}^{x+1} - q_{min}^{x+1}]_{SN1a}}~.
\end{eqnarray}
Adopting the parameters above for the carbon-oxygen white dwarf progenitor,
and progenitor masses for a low-mass black hole  in the
range of $20 \le m_1 \le 60$,  we have $M_l = 23$, $M_h = 68$, $q_{min} = 0.04$,
and $q_{max} = 0.40$ for the LMBHSN scenario.   
Setting $x = 1$ then  implies a most optimistic rate of 
\begin{equation}
{R_{LMBHSN} \over R_{SN1a}} \approx 10^{-3}~.
\end{equation}
Requiring that the white dwarf have a mass $M_{WD} > 1.2$ M$_\odot$ and that the
progenitor star produce a black hole with $M_{BH} \ge 20$ M$_\odot$ 
(as indicated in Table 1) will decrease
this rate by at least another order of magnitude.
Hence, we conclude that this type of supernova is probably much
less frequent than normal Type Ia supernovae.  The 
main reason is the simple fact that the high-mass
progenitors of a black hole are more rare than the low-mass progenitors
which can produce SNIa's.  Also, the required massive white dwarfs 
are rare,  and the required binary systems with a
high mass asymmetry are less common.  
   
\section{White Dwarfs Passing a BH}
The other plausible candidate environments for the explosion scenario
considered here are those of white dwarfs on passing orbits near a 
massive ($m_{BH} \sim 10^3$ M$_\odot$) black hole in the center of 
a globular cluster or a supermassive ($m_{BH} \sim 10^6 - 10^9$ M$_\odot$) 
black hole in a dense
galactic core.  We show here that is possible
for event rate in these environments to be greater than 
the event rate due to low-mass black-hole/white-dwarf binaries.
Supermassive black holes, for example, appear to be 
present in every galaxy.
They also tend to be found in high density stellar environments
in which a large encounter rate is possible.

\subsection{Massive Black Hole in Globular Clusters}
   At the present time there is no firm observational evidence for 
massive black holes
    in globular clusters (GCs)  even though much effort has been expended searching
    for them.  Although there have been recent claims 
(e.g. Gebhardt, Rich \& Ho 2002; van der Marel et al. 2002) 
for their detection,
it has since been demonstrated 
(cf. Baumgardt et al. 2003; Ferraro et al. 2003; McNamara 2003)
that conventional interpretations (without a massive black hole) 
can match the observations equally well. Hence there is no firm
empirical basis to conclude that most globular clusters
harbor an intermediate mass black hole.

Nevertheless, there are plausible theoretical arguments as to
why such objects should exist in at least a fraction
of globular clusters (Miller \& Hamilton 2002; Zheng 2001).
    For example, a strong correlation has been noted 
 (Gebhardt et al. 2000; Ferrarese \& Merritt 2000) 
between core black-hole masses and the dispersion
    velocity in the central bulge of the galaxies. Observed black-hole
masses scale (Tremaine et al. 2002) as,
\begin{equation}
      M_{BH} =M_0\biggl({\sigma \over \sigma_0} \biggr)^\alpha 
~~{\rm  M}_\odot~~,
\end{equation}
    where $\alpha = 4.02 \pm 0.32$ and $M_0 \approx  1.35 \times 10^8$ 
for $\sigma_0 = 200$ km s$^{-1}$.
It has been shown (Adams et al. 2003) that this scaling simply follows
because collapse to the black-hole mass is halted once 
 the centrifugal radius of the collapse flow exceeds the capture 
radius of the central black hole.
A similar collapse should account for formation of any spherical stellar
system. Hence, it is reasonable to extrapolate this relation to globular clusters.  
    For the 22 Milky-Way globular clusters 
considered in (Zheng 2001)  $\sim 10\%$ would have implied black hole masses  
    of  the order of $10^3$ M$_\odot$, but with a large uncertainty since 
    a long extrapolation is required. This fraction is also consistent
with the fraction inferred (Miller \& Hamilton 2002) from the observed
luminous point sources in several starburst galaxies.

Assuming that massive black holes exist in at least some GCs,
the rate at which white dwarfs pass sufficiently
close to a massive black hole to produce observable supernovae out to 
cosmological distances
is given by a combination of factors,
\begin{equation}
\dot N_{MBHSN} = N_{gal}\times N_{GC}\times  f_{BH}\times \lambda_{WD} ~~,
\label{rlmbh}
\end{equation}
where $N_{gal}$ is the number of galaxies out to $ z \approx  1$,
$N_{GC}$ is the average number of globular clusters per galaxy, and
$f_{BH}$ is the fraction of those clusters which contain a
black hole with $m_{BH} \ge 1000$ M$_\odot$.

For globular-cluster black holes, the rate $\lambda_{WD}$  at which white dwarfs  
pass sufficiently close to explode can be written,
\begin{equation}
\lambda_{WD} = n_{core} \langle \sigma v \rangle~~,
\label{lambdawd}\end{equation}
where $n_{core}$ is the density of  carbon-oxygen white dwarfs around the black hole.
The quantity $\langle \sigma  v \rangle$ is the average white-dwarf  gravitational
cross section times the  virial velocity distribution 
in the dense environment near the black hole. Equation (\ref{lambdawd}) assumes
an isotropic distribution of stars and a relaxation time for the system $^<_\sim$
the rate of black-hole encounters.

The gravitational  cross section $\sigma$ can be written in terms of a maximum
asymptotic impact parameter $b$,
\begin{equation}
\sigma = \pi b^2 ~~,
\end{equation}
such that  a white dwarf
passes  close enough to the  black hole for its density to exceed the
pycnonuclear  ignition threshold.

First, let us estimate the cross section $\sigma$.
A white dwarf infalling from a large distance will have a
specific  angular momentum:
\begin{equation}
\tilde L = \tilde E v_\infty b~~,
\end{equation}
where  $v_\infty$ is the  tangential velocity component of the white
dwarf at a large distance from the black hole.
The specific mass-energy is given by
$\tilde E \approx 1.0 + v_\infty^2/2$~~.
This gives
\begin{equation}
\tilde L = \tilde E  b \sqrt{ 2 (\tilde E - 1)}~~.
\end{equation}
Now, at closest approach $z_{min}$, 
\begin{equation}
\tilde E = \sqrt{(1 - 2M/z_{min})(1 + \tilde L^2/z_{min}^2)}~~,
\end{equation}
which can be solved for the impact parameter $b$ to give:
\begin{equation}
b^2 = { 2 M z_{min} \over  v_\infty^2( 1 - 2M/z_{min})}~~.
\end{equation}
Now inserting,
$U \approx \dot \theta z_{min} = \tilde L/z_{min}$~~,
leads to
\begin{equation}
U^2 \approx  b^2 v_\infty^2/z_{min}^2 =   { 2 M \over z_{min} ( 1 - 2M/z_{min})}~~.
\end{equation}
For the cases of interest in Tables 2 and 3, $M/z_{min}$ is small at the distance 
at which the ignition density is reached and can be ignored.
Solving this for $b^2$ 
and substituting $U^2 \approx  2 M/z_{min}$,
gives the desired cross section as a function
of $v_\infty$ for the white dwarf, $M$ for the black hole,
and the $U^2$ at which the star can explode,
\begin{equation}
\sigma = \pi b^2 = {4 \pi M^2 \over v_\infty^2 U^2 }~~.
\label{sigma}
\end{equation}
Equivalently, one can write the cross section in terms of the
distance of closest approach, 
\begin{equation}
\sigma =  {2 \pi M z_{min} \over v_\infty^2  } 
= \pi z_{min}^2 \biggl({U^2 \over v_\infty^2  }\biggr)
\label{sigma2}
\end{equation}

Let us consider a typical globular cluster (Binney \& Tremaine 1987)
 with a core radius of 1.5 pc,
a core density of $n_{core} \approx 8 \times 10^3$ M$_\odot$ pc$^3$, 
and a core virial velocity
of $v_\infty \approx 7 $ km s$^{-1}$.
Even at high redshift, the core is likely to be dominated by white dwarfs
with a mass distribution peaked around 0.6 M$_\odot$ but with a broad
distribution extending up to 1.2 M$_\odot$ (Silvestri et al. 2001).
We have determined the rate at which  supernovae of this type are ignited 
per globular cluster by integrating Eq.~(\ref{lambdawd}) over 
a thermal distribution of white dwarf velocities normalized
to a virial velocity of $v_\infty \approx 7 $ km s$^{-1}$. The velocity dependent 
gravitational cross section was taken from Eq.~(\ref{sigma2}) with an 
average for $z_{min}$ taken from the results of Table 2 weighted by the
mass distribution of Silvestri et al. (2001). We determine in this way a rate of
$\lambda_{WD} = 2 \times 10^{-8}$ SNe per year per massive black hole in a globular cluster.
Hence, the use of Eq.~(\ref{lambdawd} is justified
since our estimated black-hole encounter
rate is less than  the median core relaxation time derived for 
GCs in the Milky Way (Binney \& Tremaine 1987).

The number of galaxies out to a redshift $z \approx 1$ can be obtained by integrating the 
galaxy luminosity function and is approximately $N_{gal} \approx 10^9$.
The number of globular clusters per galaxy  is a strong
 function of the parent galaxy luminosity
(Vandenbergh 1997).  For example, in $M87$ there are $ 1.4 \times 10^4$ globular
clusters while the Milky Way only contains about 150.
Since globular clusters are destroyed over time, it is quite possible the
the low number of GCs in the Milky Way is atypical.
If we take the number of GCs around $M87$ as more realistic,
 then there could be more than $ 10^{13}$
globular clusters out to a redshift of $z \approx 1$.  

As of yet there are no confirmed 
globular clusters which contain a massive black hole
in the Milky Way.  Nevertheless,  as argued above it is at least possible
 that a fraction of GC's contain an  as yet undiscovered
massive black hole. Therefore, we allow  for the possibility that $f_{BH} >  0$.  
The whole sky rate for this class of supernova then becomes 
\begin{equation}
\dot N_{MBHSN} \approx 10^5 {N_{gal} \over 10^9} {N_{GC} \over 10^4} f_{BH} ~~{\rm yr}^{-1}~~.
\end{equation}
Although this estimate (like that for LMBHSNs) is smaller than the observed  ($\sim 10^7$ yr$^{-1}$)
Type Ia supernova rate out to large redshift, it could be much greater if for example,
the average number of globular clusters per galaxy were larger in the past. 
Hence, even for $f_{BH}~^<_\sim 0.1$, we 
conclude that this class of supernova event may be detectable.

\subsection{Supermassive Black Holes in Galactic Centers }
We next  estimate the rate
for the explosion of white dwarfs on passing orbits
near supermassive black holes in a galactic centers.
Such systems may be natural candidates for the 
events described here.  For one,  it appears that almost any moderate size
galaxy has a supermassive black hole at its core (Kormendy \& Richstone 1995;
Magorrian et al. 1998) with a mass ranging from
$10^6$ to $10^9$  M$_\odot$ (Gebhardt et al. 2003).  
Furthermore, it is well know that quasars and AGN's are
absorbing of order 1 to 10 M$_\odot$ of material per year,
and that there is a very high stellar concentration around the central
supermassive black hole.  For example the modest $3 \times 10^6$  M$_\odot$
SgrA$^*$ black hole in the Milky Way has an observed mass concentration 
after subtracting the black hole
(Ghez et al. 1998; Genzel et al. 2003) of $1.2 \times 10^6 \times (R/10")$  M$_\odot$ pc$^{-3}$
where R is the angular distance from SgrA$^*$ in arcseconds 
(1 pc $\approx$ 24" in the Galactic center).
In the inner cusp, the density can exceed $10^8 $  M$_\odot$ pc$^{-3}$.
Although many of these stars appear to be members of
a young stellar population (Figer 2000; Gezari et al. 2002), it seems likely that
these young stars are the result of mergers from
older stars (Sch\"odel et al. 2003).  Hence, a comparable population of
white dwarfs could be present.  We conservatively estimate
$n_{wd} \sim 10^6$   pc$^{-3}$  around the black hole
with an average velocity dispersion of
$v_\infty \approx 100$ km s$^{-1}$.

In the case of a galactic center, equation (\ref{lambdawd})  cannot be used to
estimate the steady state rate at which stars fall into the BH. 
This is because stars
on eccentric orbits that bring them close enough to the BH to be
destroyed ("loss-cone" orbits) are depleted rapidly on a dynamical
time scale.  Subsequently, the rate is set by the much slower
diffusion time-scale of stars from BH-avoiding orbits into radial
orbits.   Here we will apply the loss cone model as discussed in 
Syer \& Ulmer (1999) and Magorrian \& Tremaine (1999).

In this model the two-body relaxation timescale for 
a spherical cluster of stars of mass $m_*$
with a density $\rho(r)$ and an isotropic density dispersion $\sigma(r)$
is given by,
\begin{equation}
t_r = {\sigma^3\over \ln{\Lambda} G^2\rho m_*},
\label{tr}
\end{equation}
where $m_*$ is the typical stellar mass and
$\Lambda$ includes dimensionless factors of order unity as well
as the Coulomb logarithm. Following Binney \& Tremaine (1987) and
Syer \& Ulmer (1999) we then can write a numerical value of:
\begin{eqnarray}
t_r &=& \frac{1.8\times 10^{10} {\rm~y}}{\ln(0.4 \times N) }
\biggl({\sigma \over 100 {\rm~km~s^{-1}}}\biggr)^3
\nonumber \\
&& \times 
\biggl({M_\odot\over m_*}\biggr) \biggl({10^4 M_\odot 
{\rm~pc^{-3}}\over\rho}\biggr) ~~,
\label{trnum}
\end{eqnarray}
where $N$ is the number of stars within the characteristic radius
$r_b$ and $\sigma$ is the one-dimensional velocity dispersion
(Seyer \& Ulmer 1999) given as
\begin{equation}
\sigma^2 \approx \frac{GM}{(1 + p)r}~~,
\end{equation}
where $p \sim 2$ is the logarithmic slope of the density. 
The white dwarf capture rate will then be given by the
rate at which white dwarfs will scatter into the loss cone.
\begin{equation}
\lambda_{WD} \approx  \int \frac{4 \pi n_{wd}(r)  r^2 dr}
{\ln{(2/ \theta_{l c})} t_r}~~,
\end{equation}
where $n_{wd}(r)$ is the number density of white dwarfs
\begin{equation}
n_{wd} = f_{wd} \rho/\langle m_{wd} \rangle~~,
\end{equation}
where $ f_{wd} \sim 0.5 $ is the mass fraction in white dwarfs and 
$\langle m_{wd} \rangle \sim 0.6$M$_{\odot}$ is the mean white-dwarf mass. 
The quantity $\theta_{l c} = q G M/r^2 \sigma^2$ is the angular size
of the "loss cone" where $q$ is the radius  at which 
stars are removed from the system. 
The density profile $\rho(r)$  divides into inner and outer regions.
The black hole dominates the gravity in the inner region so that a density
profile of the form $\rho \approx r^{-2}$  and 
a dispersion velocity, $\sigma^2 \approx  GM/3r$ is expected.
In the outer region on the other hand one expects flatter density 
and velocity-dispersion
contours.  Based upon the analysis of Syer \& Ulmer (1999)
(cf. their figure 4) we estimate
$\lambda_{WD} \sim  3 \times 10^{-5}$
for a typical white dwarf mass of $0.6$ M$_\odot$.  
However, based upon the range and uncertainties in the
parameters $\rho$, $\sigma$, $r_b$, and M$_{BH}$
we expect a range in $\lambda_{WD}$ 
from up to an order of magnitude above to
a couple of orders of magnitude below this number.
For purposes of discussion we will adopt $\sim  3 \times 10^{-5}$
as a reasonable estimate.

The total rate of such events out to $z \approx 1$ 
will be given by
 a similar combination of the factors
to those  given above in Eq. (\ref{rlmbh}),
\begin{equation}
\dot R_{SMBHSN} = N_{gal} N_{SM}  \lambda_{WD} ~~,
\end{equation}
where again, $N_{gal} \sim 10^9$ is the number of galaxies out to $ z \approx  1$ and 
$N_{SM} \sim 1$ is the number of supermassive black holes per galaxy. 
The quantity $\lambda_{WD}$ is again the rate at which white dwarfs are making
sufficiently close passes to a black hole.  The total rate is then of order
10$^3$ to  a few $\times10^5$ yr$^{-1}$.  
This rate is somewhat less than normal SN1a's which are
observed at a rate $\sim 10^7$ yr$^{-1}$. 
Furthermore, a large fraction of such events
might be hidden in the dense regions surrounding galactic cores.
Nevertheless, this  crude estimate is at least suggestive
that such events could be frequent enough to be observed.
It is conceivable that they might
even dominate  over normal Type Ia supernovae at sufficiently large redshift 
where newly formed galactic cores  have not yet depleted their
radial orbits.  Frequent white dwarf encounters with a supermassive black
hole might even provide a power source for AGNs  and quasars.
Clearly, if such explosions occur, a search for observed signatures 
seems warranted. 

\section{Possible Observed Signatures}
Having proposed this class of new supernovae it is worth
commenting on several distinguishing 
characteristics which might help observers identify
whether one of these events has  actually occurred. 
Preliminary numerical burn simulations (D. Dearborn, priv. comm.)
of 0.6 M$_\odot$ white dwarfs passing a supermassive black hole indicate
a nearly complete burn of initial C/O into Ni/Co.  Hence,
each supernova  should  indeed generate $\sim 10^{51}$ erg in kinetic energy of matter
and $\sim 10^{49}$ erg in optical light.  
The optical signal, however, may be unique for several reasons.

 One obvious feature is that these supernovae should most likely be seen 
in association with a galactic core (or possibly a globular
cluster).  Indeed, a number of transients that
might be such supernovae near galactic cores
have been observed (N. Suntzeff, priv. comm.)
out to large redshift, but are often not analyzed
due to the complication of their location in a crowded field.
We suggest that in
the near future, the development of the  National Virtual Observatory
(cf. Hanisch 2002)
might  be utilized to search for a possible  association of unusual Type I supernovae
with galactic cores and/or  extragalactic globular clusters.

Among other
things to look for is the fact that the stars are moving very fast when they explode. 
Hence,  they could exhibit a large
Doppler shift in their spectra.
Also, the explosion itself should be different, 
since the stars are of lighter mass and undergo  a very rapid
rapid compression/ignition.  Normal Type Ia events are expected
to undergo an off-center ignition after a gradual approach to the
ignition density. The events described here, however, will probably ignite rapidly and
at their centers.  This will cause a detonation and a possibly  more efficient
thermonuclear burn of the white dwarf (depending upon tidal effects from
the black hole).  Thus, one might expect a different burn front,  a different light curve,
and different nucleosynthesis yields (e.g.~more iron and reduced silicon) 
compared to normal SNIa events.

\subsection{Interaction with the Black Hole}
Another distinguishing characteristic of these explosions is that they
occur in the strong field of a companion black hole.  Hence, there  might 
be an affect on the light curve from matter accreting into the black hole
and/or impinging on the accretion disk.
For example, the accretion of SN debris onto the black hole
might produce excess X-ray (or even $\gamma$-ray) emission.

 We suggest two possible observational
consequences of the SMBHSNe, depending upon the optical depth of the
material in the accretion disk down to the location where the supernova occurs.  
    If the optical depth to the point of explosion is greater than one, then since
    the scale of the systems is so large,  energy will be emitted from the system
over  a timescale  of a year or more. 
    
In the case that the optical depth is less than one then the usual SN light signal
    will be seen. However, if there is an accretion disk  present, 
then the SN light signal will be augmented
    by radiation emitted from the collision of the SN remnant with the accretion disk.

    Let us consider order-of-magnitude estimates for the radiation emitted as debris 
from the explosion of a 0.6 M$_\odot$
    white dwarf impinges on the accretion disk of a $10^9$ solar mass black hole. For this we will
assume that the accretion disk is flat.  That is, we ignore the 
thickness of the accretion disk near the black hole.

  From Table 3, the distance from the black hole  to the point at which the explosion 
initiates is 
typically about $d=1.5\times 10^{15}$ cm (roughly ten black hole Schwarzschild  radii).
We take this as a measure of the distance from the supernova to the
accretion disk.  
 The expansion velocity of the supernova shell is given by 
$v \approx  (2\times 10^{51}/{\rm M}_{WD})^{1/2} = 1.3\times 10^9$ cm s$^{-1}$. 
Hence, the time scale  for material to reach the accretion disk is of the order  
$1.5\times 10^{15}/v \approx   10^6$ sec.
    The time for peak luminosity of a Type-I supernova is also about $10^6$ sec. Thus, the 
    peak supernova signal and the start of the thermal radiation from the collision of the 
supernova remnant with the accretion disk will
typically occur at about the same time. 

The peak temperature from the collision shock will be about
    400 keV. The electrons and baryons will come to equilibrium  temperatures quickly. Then
    bremsstrahlung radiation will cool the hot gas on a timescale  of about $10^4$ sec. Most of
    the remnant will not hit the accretion disk perpendicularly. Averaging the various
    physical quantities over the angle of impact we estimate a mean photon energy of
    about 30 keV. The time for the photon spectrum to fall below 10 keV will be about
    $10^7$ sec. The total emitted X-ray energy will be $\sim 10^{50}$ ergs. Thus, an X-ray
    burst in coincidence with a supernova explosion deep inside a galaxy is what
    we expect to possibly be observed.

\subsection{Remnants}
  As another observed signature, it is possible that remnants from this
type of supernova may be detectable in the Galaxy.
For example, an event rate  of a few  $\times 10^{-5}$ y$^{-1}$ per galaxy  
for white dwarfs to approach and explode near the 
giant black hole in galaxy centers, suggests that it is worthwhile
to search for a supernova remnant with an age $\approx 3 \times  10^4$ y
near the giant black hole SgrA* in the
Galactic center.  For a remnant moving with a 
typical dispersion velocity of 100 km s$^{-1}$
for $10^4$ y, one would expect to find such a remnant $\sim$ 1 pc
from SgrA*.  

In this regard, it is certainly of interest
that indeed such a remnant (SgrA East) exists and has been recently analyzed
(Maeda et al. 2002) with the ACIS detector aboard the {\it Chandra X-ray Observatory}.
This remnant is centered at a distance of $\approx 3$ pc from SgrA*.  In fact,
SgrA* is within the remnant outer rim.  The passage of the  supernova 
shell  sometime in the past may have swept gas away from the black-hole vicinity and
hence, may  be responsible for the present quiescent state of SgrA*.
  Also of note is the fact that its peculiar 
metal abundances,  along  with the unusual combined radio and X-ray morphologies classifies
this remnant as a metal-rich "mixed-morphology" (MM)  SNR.  Although, the
properties of SgrA East  may be explained  (Maeda et al. 2001)
by a low-mass Type II supernova, it is tempting here to speculate
that this class of MM SNRs may be the smoking gun for a 
black-hole induced white-dwarf explosion (cf.~Khokhlov \& Melia 1996).

Obviously, a clear picture of how to classify these events will
require fully relativistic numerical computations 
in three spatial dimensions of the compression and 
 thermonuclear explosion of a   white dwarf
in the background field of a black hole.
Efforts along this line are currently under way
(D. Dearborn priv. comm.) utilizing three dimensional hydrodynamics
and radiation transport.  As noted above, preliminary results indicate
a central thermonuclear ignition (very close to our adopted  
ignition density) which proceeds to detonate the star.
In future work we will simulate the radiation transport,  
light curve, and supernova remnant.

\section{Conclusions}

  We have explored a new class of Type-I supernovae whereby the onset of 
a thermonuclear explosion is induced by relativistic enhancements of the
white-dwarf self gravity as it accelerates in the background gravitational field
of a black hole.  We find that a potentially observable rate of such events
could be occurring out to cosmological distances, particularly 
in the dense regions of galactic cores and possibly globular clusters as well.  
These explosions and their remnants could thus be characterized
by their association with supermassive black holes galactic cores or
 massive black holes in globular clusters.
They  also might be characterized by
an associated x-ray burst 
and/or a significantly modified light curve
from matter colliding with the black-hole accretion disk.

\acknowledgments
Work at the Lawrence Livermore National Laboratory performed in part
under the auspices of the U.~S.~Department of Energy under contract
W-7405-ENG-48 and NSF grant PHY-9401636.
Work at the University of Notre Dame
supported  by the US Department of Energy under Nuclear Theory grant
DE-FG02-95ER40934.


\end{document}